\def\lea{\mathrel{<\kern-1.0em\lower0.9ex\hbox{$\sim$}}}
\def\gea{\mathrel{>\kern-1.0em\lower0.9ex\hbox{$\sim$}}}

\documentclass[preprint]{aastex}

\slugcomment{Accepted by The Astronomical Journal}

\shorttitle{M31 Bulge MDF}
\shortauthors{Sarajedini \& Jablonka}

\begin{document}


\title{The Metallicity Distribution Function of \\ Field Stars in
  M31's Bulge}

\author{Ata Sarajedini}
\affil{Department of Astronomy, University of Florida, Gainesville, FL 32611}

\author{Pascale Jablonka\footnote{on leave from GEPI CNRS-UMR~8111, Observatoire de Paris, 5 Place Jules Janssen, F-92195 Meudon Cedex, France}}
\affil{Observatoire de Gen\`eve, Laboratoire d'Astrophysique de l'Ecole Polytechnique
Federale de Lausanne (EPFL),  CH-1290 Sauverny, Switzerland}



\begin{abstract}
We have used Hubble Space Telescope Wide Field Planetary
Camera 2 observations to construct a color-magnitude
diagram (CMD) for the bulge of M31 at a location $\sim$1.6 kpc from the 
galaxy's center.
Using scaled-solar abundance theoretical red giant branches with a range of 
metallicities, we
have translated the observed colors of the stars in the CMD to abundances
and constructed a metallicity 
distribution function (MDF) for this region. The MDF shows a peak at
$[M/H]$$\sim$0 with a steep decline at higher metallicities and a more
gradual tail to lower metallicities. This is similar in shape to the MDF
of the Milky Way bulge but shifted to higher metallicities by $\sim$0.1
dex. 
As is the case with the Milky Way bulge MDF,
a pure closed box model of chemical evolution, even with significant 
pre-enrichment, appears to be inconsistent with the M31 bulge MDF.
However, a scenario in which an initial infall of gas enriched the
bulge to an abundance of $[M/H]$$\sim$--1.6 
with subsequent evolution
proceeding as a closed box provides a better fit to the observed MDF.
The similarity between the MDF of the M31 bulge and that of the Milky Way 
stands in stark contrast to the significant differences in the MDFs of 
their halo populations. This suggests that the bulk of the stars in the bulges of 
both galaxies were in place before
the accretion events that occurred in the halos could influence them. 
\end{abstract}




\section{Introduction}

In order to probe the star formation histories of nearby galaxies,
it is vitally important to resolve their stellar populations into
individual stars. The color-magnitude diagram (CMD) 
is particularly powerful in this regard because it allows the
determination of astrophysically interesting properties
from specific CMD features. One example is the luminosity of the
main sequence turnoff, which is sensitive to the age of the population.
Another example is the mean dereddened color and color dispersion of 
the red giant branch (RGB) which is a sensitive probe of the global 
metal abundance and its dispersion (detailed chemical species remain 
the art of spectroscopy). 

Given that the most propitious manner in which to study
stellar populations is to resolve them into individual stars, it is
not surprising that so little is known about the nature of spiral
galaxy bulges. Not only does the incredible crowding make it difficult
to resolve them into stars, but there are only two examples of these
systems in the Local Group - the Milky Way (MW) and M31 bulges.  In the
case of the latter, most of the recent ground-based studies have been
conducted in the near-infrared JHK passbands where advances in
adaptive optics (AO) have allowed high resolution imaging to be
performed.  While initial studies (e.g. Davidge 2001) focused on
the nature of the asymptotic giant branch (AGB) stars, the advent of
8-10m telescopes equipped with AO has provided the capability to study
individual M31 bulge RGB stars from the ground. Davidge et al. (2005)
have used the Gemini-North telescope with the ALTAIR natural guide
star AO system to image two M31 bulge fields (2' and 4' from the
center) with a FWHM of 0.08" in the K' filter. They find that the
color of the RGB stars in the inner field suggests a metal abundance
consistent with that of the Galactic globular cluster NGC 6528
($[Fe/H]$$\sim$0). An intrinsic dispersion of $\pm$0.5 dex in
metallicity is also seen.

In contrast with the M31 bulge, ground-based studies of the M31 halo
are relatively ubiquitous in the literature even at optical wavelengths.
This is because the reduced level of crowding in the halo and the larger 
fields of view available with ground-based optical imagers combine to produce an
excellent match thus yielding impressive results. For example,
Ferguson et al. (2002) have conducted a large survey of the M31 halo with
the Isaac Newton Telescope Wide Field Camera. They report on the existence
of  significant stellar substructures in the M31 halo. More specifically, 
they find evidence for spatial density and metallicity
variations and conclude that the differences between the stellar halos of
M31 and the MW - for which substructures are also found at large
radii (Newberg et al. 2002) - may be the result of a more active
accretion history for the former.

Space-based optical and near-IR observations such as those with the
(refurbished) Hubble Space Telescope (HST) have also been utilized in
studies of M31's bulge. For example, Jablonka et al. (1999) present
HST observations of three fields in the region of the M31 bulge taken 
with the Wide Field Planetary Camera 2 (WFPC2). They showed that their
photometry for stars on the WF chips was severely comprised by the
extreme crowding present in the bulge of M31. However, the photometry
for stars on the PC1 chip was relatively robust and was therefore used
to show that, contrary to the results of Rich \& Mighell (1995), the
tip of the first ascent RGB in M31 does not become significantly more
luminous ($\gea1$ mag) inside of $\sim$2 kpc of that galaxy's center.
Stephens et al. (2003) have used HST as well to image the bulge of
M31, but they utilized the NICMOS instrument in the J, H, and K
passbands. Their primary conclusion is that the luminosity function of
RGB stars in the M31 bulge is not measurably different from the bulge
of the MW.

HST/WFPC2 investigations of the stellar populations in M31's disk
and/or halo 
include those of Williams (2002), Bellazzini et al. (2003), Brown et al.
(2003), and Renda et al. (2005). Brown et al. (2003), which is
especially relevant in the present context, is the first study to reach
the old main sequence turnoff in the halo of M31. They conclude that
the stellar population is composed of an old and metal-poor component along
with an intermediate-age, more metal-rich population.

Thus, while it seems that M31 and the Galaxy differ in a number of
important ways (e.g. Renda et al. 2005), it is still not clear whether
we can describe their formation and evolution within a common theoretical 
framework. An important piece of the puzzle
that is still missing is a more complete understanding of the
properties of M31's central spheroid.  This leads us to the principal
aim of the present paper which is to derive the metallicity
distribution of the bulge of M31.  This is the first determination of
the MDF for the bulge of an external galaxy.

We start with the Jablonka et al. (1999) HST/WFPC2 dataset and use it
to probe the metallicity distribution function (MDF) of the M31 bulge
stellar population. The next section describes the observational data.
Section 3 discusses the MDF and how it is constructed. A comparison of
the MDF with that of the Milky Way and an analysis in terms of
chemical enrichment models is presented in Sec. 4.

\section{Observational Data}

The photometry analyzed in the present work is that of Jablonka et al.
(1999, hereafter J1999) who observed the fields around three M31 bulge
clusters - G170, G198, and G177 - with the HST/WFPC2 instrument 
(GO-5907 and GO-6477). They imaged the G170 and G198 fields for a total 
exposure time of 5200s in F555W and 4800s in F814W with a 2-point
dither pattern. The G177 field, being closer to the center of M31
(see below), was observed 
for a total of 7800s in F555W and 7400s in F814W with a 3-point
dither pattern.

The clusters G170, G177 and G198 are located at 6.1(1.55 kpc SW),
3.2(0.80 kpc SW) and 3.7(0.97 kpc NE) arcmin from the center of M31,
respectively (Hodge 1981). Bulge decompositions show that
the surface brightness of the bulge
at these galactocentric distances is at least two magnitudes brighter
than that of the M31 disk (Walterbos
\& Kennicutt 1988 ;  Kent 1989). Therefore, we can be confident that the
regions around G170, G177, and G198 predominantly sample the bulge of
M31.

J1999 constructed VI color-magnitude diagrams (CMD) for these fields
and studied the variation of the red giant branch (RGB) tip magnitude
with radial position in the M31 bulge. Using artificial star
experiments, J1999 showed that the stellar crowding in these fields
was so severe that it compromised the photometry derived from the WF
chips so that only the PC1 observations were usable. These are the
data we will analyze herein.

The reduction of the PC1 images was fully described by J1999. Here,
we provide a brief summary of the procedure.  The images were 
photometered with the DAOPHOT/ALLSTAR/ALLFRAME PSF-fitting
package (Stetson 1987;1994). We used the high signal-to-noise empirical 
PSFs constructed for the HST Cepheid distance scale key project (Stetson 
1996) and the accompanying photometric transformation of 
Silbermann et al. (1996).  Keeping in mind the extreme crowding of these M31 
bulge images and that the dither pattern of the observations 
was designed to optimize the ALLFRAME reductions, we are
convinced that a new reduction of these images will not yield 
significantly improved photometry. Therefore, we proceed to analyze the
J1999 data as it was originally reduced.

Our first task is to better understand the photometric completeness of
the three bulge field datasets. If we are to use the color
distribution of the RGB stars to construct a MDF, then we must assure
good completeness for the RGB stars over their entire color
range. Figure 1 shows the CMDs of the G170, G198, and G177 fields
while Fig. 2 shows the corresponding I-band luminosity functions (LF).
Each of the fields displays a prominent RGB extending from I$\sim$21
to the limits of the photometry. In addition, the G170 field also
shows the helium-burning red clump stars at I$\sim$24.5.

Given the power-law form of the LF along the RGB (Girardi et al. 2002), Fig. 2
suggests that the G177 and G198 fields suffer from photometric
incompleteness at magnitudes below V$\sim$24.5 mag , I$\sim$23.5 mag.
This in turn implies that the photometry in these fields is probably not
adequate to allow the construction of a robust and
meaningful MDF. In contrast, the data in the G170 field appears to be
complete to below the red clump at V$\sim$26 mag, I$\sim$25 mag. This
is corroborated by the artificial star experiments performed by
J1999. They found a 91\% completeness rate for RGB stars as faint as
I=25. Based on these results, we proceed to utilize the RGB stars in
the G170 field to construct a MDF of the M31 bulge.

\section{Metallicity Distribution Function}

Figure 3 shows our G170 field data adjusted for the distance and reddening of M31.
To remain consistent with our previous work, we have adopted 
$(m-M)_0 = 24.43$ (Freedman \& Madore 1990) and $E(B-V) = 0.062$
(Schlegel, Finkbeiner, \& Davis 1998) leading to $(m-M)_I = 24.55$ and
$E(V-I) = 0.08$. The solid lines are the RGB sequences from the theoretical
isochrones of Girardi et al. (2002) for an age of 12.6 Gyr and metallicities of
Z=0.0001, 0.0004, 0.001, 0.004, 0.008, 0.019, 0.04, and 0.07 for a scaled-solar
abundance mixture. Based on 
a suggestion from Leo Girardi, we have combined the models of Girardi et 
al. (2002) for metallicities up to Z=0.019 with those of Salasnich et al.
(2000) for higher metallicities. Thus, our model grid represents a self-consistent 
set of theoretical RGBs with a range of abundances.

In order to use the RGBs in Fig. 3 to calculate metallicities for each
star, we must adopt a mean age for this region of the M31
bulge. Citing a number of previous results, Rich (2004) argues that,
like the MW, the M31 bulge is likely to be old (i.e. $\gea$10
Gyr). 
Under this assumption, we adopt a mean age of 12.6 Gyr and note
that the precise value of the age is not important because of the
insensitivity of the RGB colors to age for old stellar populations.
Tests we have performed show that an age change of $\pm$0.1 in the
logarithm from our nominal 10.1 value (10 Gyr $\leq$ Age $\leq$ 15.8
Gyr) results in a change of less than $\pm$0.05 dex in the peak metal
abundance of the population.

A number of studies have found that stars in the Milky Way bulge are
enhanced in the $\alpha$-capture elements such as oxygen, magnesium,
and silicon (e.g. McWilliam \& Rich 2004).
There are hints that the M31 bulge stars are also enhanced in
$\alpha$-elements, as suggested by the spectroscopic analysis of its
nucleus (Worthey, Faber \& Gonz\`alez, 1992), though no definitive
value is known. We are deriving metallicities from RGB
stars, and the position of the RGB is mainly a function of Si and Mg,
as far as the alpha-elements are concerned (Renzini, 1977). Since
no theoretical isochrones are presently available with differential
elemental abundances, we will estimate
the effect of such an enhancement on our MDF analysis using the
formalism of Ferraro et al. (2000) wherein
\[
[M/H] = [Fe/H] + log(0.638f_{\alpha} + 0.362)
 \]
with $f_{\alpha}$ representing the factor by which the $\alpha$-elements
are enhanced. Thus, if $[\alpha/Fe]$ = +0.3, then $f_{\alpha}$=2 and
our computed scaled-solar metallicities will need to be increased by
$\sim$0.2 dex to account for this $\alpha$ enhancement.

We now proceed to use the $M_I$ and $(V-I)_0$ values of the RGB stars
to interpolate among the theoretical sequences of known abundance
(Fig. 3). The isochrones represent a grid indicating the locations of
stars with particular values of $M_I$, $(V-I)_0$, and [M/H]. From the
colors and magnitudes of the RGB stars, we can infer their metallicities
using this theoretical grid. We have utilized the `TRIGRID' routine in
the Interactive Data Language (IDL) to perform the interpolation. TRIGRID
uses the method of `Delaunay triangulation' to set up a three-dimensional 
grid of regularly spaced points
within which a polynomial is used to interpolate. The error in each
metallicity is determined by calculating the uncertainty due to errors
in $M_I$ and $(V-I)_0$ individually and adding these in quadrature.
Our nominal set of RGB stars includes those with $M_I$$\leq$$-1.2$ 
(so that we exclude the red clump stars) and contains 7771 stars. 

Figure 4 displays the binned (filled circles) and generalized
histograms representing the metallicity distribution function (MDF) of
the M31 bulge region near the cluster G170. The latter has been
constructed by adding up unit Gaussians with widths given by the error
in each metalliciity. The generalized MDF has at least two advantages
over the binned one. First, it is not subject to the vagaries
associated with binning - the choice of endpoints and the bin
sizes. Second, genuine (and artifactual) features of the distribution
are revealed when the errors are taken into account in its
construction.  Figure 4 shows that the M31 bulge MDF peaks at
$[M/H]$$\sim$0.0 with a steeper dropoff to the metal-rich end as
compared with the metal-poor region. 

We close this section by reiterating that the MDFs shown in Fig. 4
are largely insensitive to the age of the isochrones as long as
an age older than $\sim$10 Gyr is adopted. In addition, we explored the
effects of errors in our reddening and distance modulus values on the shape
of the resultant MDFs.  Errors of $\pm$0.02 in $E(V-I)$ and $\pm$0.1 mag
in $(m-M)_I$ produce MDFs that have the same peak abundance
($[M/H]$$\sim$0) and very similar metal-rich and metal-poor tails
suggesting a negligible effect on our conclusions. Furthermore, MDFs were
constructed using stars in two additional magnitude ranges 
(--1.2 $\geq$ $M_I$ $\geq$ --1.7 and --1.2 $\geq$ $M_I$ $\geq$ --2.2). Once
again, no significant differences were seen in the properties of the distributions.

\section{Results and Discussion}

We first compare the M31 bulge MDF with that of the Milky Way's bulge.
The most recent and arguably most comprehensive such MDF has been
produced by Zoccali et al. (2003, hereafter Z2003) who combined optical and
near-infrared photometry for stars toward the Galactic bulge. Relying
on a technique similar to our's, Z2003 converted their
photometric colors to metallicities using fiducial globular cluster
V--K RGBs covering the range $-1.9\leq[M/H]\leq-0.1$. 
In order to be consistent with our M31 bulge MDF, we have rederived the Milky
Way MDF using the Z2003 V--K photometry coupled with the scaled-solar
Girardi et al. (2002) and Salasnich et al. (2000) model RGBs for the same 
metallicities as those shown in Fig. 3. Figure 5a displays these theoretical
RGBs in V--K along with the Z2003 data.  It is important to note
that the original Z2003 Milky Way bulge MDF incorrectly included a number of
stars above the first ascent RGB tip (i.e. AGB stars), which we have not 
included here. The resultant MDF based on 471 MW bulge stars and
corrected for the effects illustrated in Fig. 13 of Z2003,  is shown in the
bottom panel of Fig. 5 and compared with our
M31 bulge MDF in Fig. 6. Both histograms have been scaled to unit
area.


We see that both the MW and M31 field bulge populations have a peak
abundance close to the solar value, with the M31 peak being at most
0.1 dex more metal-rich. However, the MW MDF appears to be slightly 
more peaked as
compared with that of M31.  In addition, the low metallicity tail and
the high metallicity drop-off of the M31 distribution both seem to be
shifted to higher [M/H] values by 0.1 to 0.3 dex relative to the MW.
As far as the $\alpha$-elements are concerned, if both the MW and M31 bulge 
populations are enhanced by the same amount, both MDFs with shift accordingly
leaving the difference between them unchanged. However, if there
is a difference of $\pm$0.3 dex in  $[\alpha/Fe]$, the MDFs will shift by
$\pm$0.2 dex in [M/H] relative to each other.
It therefore seems that, assuming the abundance of the $\alpha$-elements is 
comparable, the M31 bulge MDF and that of the MW are 
remarkably similar in shape with the former being slightly more
metal-rich than the latter.

Following Z2003, we can analyze
the M31 bulge MDF in the context of the closed-box chemical enrichment
model (Pagel 1997).  This simplest of all enrichment scenarios assumes
that no infall or outflow of gas occurs during the star formation
history of the stellar system. It typically also adopts the
instantaneous recycling approximation, which asserts that processes
such as stellar evolution, nucleosynthesis, and elemental recycling
take place on a timescale much shorter than the age of the system.
This approximation is generally sound for a system that is old
($\sim$10 Gyr) regardless of the precise element being
studied. 

According to Pagel (1997), the number of stars at each
abundance is given by
\[
\frac{dN}{dZ}\sim \frac{1}{y}e^{-(Z-Z_0)/y}~~~~~~~~~
\frac{dN}{d[M/H]} \sim \frac{Z-Z_0}{y} e^{-(Z-Z_0)/y}
\]
In this formulation, $Z_0$ is the initial metal abundance of the gas.
The parameter $y$ is known as the yield and is equal to the mean metal
abundance of the system. Figure 7 shows a pure closed box model
prediction ($Z_0 = 0$) with $y=$$\langle$$Z$$\rangle$=0.019 [the solar
abundance in the Girardi et al. (2000) models] along with our M31
bulge MDF. The dashed line in the upper panel of Fig. 7 illustrates
the comparison using the logarithmic quantity $[M/H]$ while the lower
panel displays the result of performing the comparison in the linear
abundance plane. Note that the [M/H] MDF has been normalized to fit
the peak of the distribution while the $Z/Z_{\odot}$ MDF has been
normalized to the metal-rich tail. We see that the closed-box model
provides an excellent fit to the region around the peak of the MDF in
the upper panel. However, more metal-poor than $[M/H]$$\sim$--0.8, the
model consistently predicts more stars than are present in the
observations. This feature is indicative of the classic G-dwarf
problem (van den Bergh 1962; Schmidt 1963) for which a number of
solutions have been proposed.

One solution to the G-dwarf problem postulates a nonzero initial
abundance for the gas in the closed box that formed the first
generation of stars. In this case, $Z_0$$>$0 in the equations
above. As an example, the dashed lines in Fig. 8 show a closed box
model with $Z_0 = 0.0005$ ($[M/H] = -1.6$). It is clear that while the
pre-enriched closed box scenario fits the logarithmic MDF in the upper
panel of Fig. 8, it does a poor job of matching the low metallicity
tail of the linear MDF in the lower panel. Interestingly, Harris \& Harris
(2000) also noticed a similarly poor fit at low Z, albeit in the halo of the peculiar
elliptical galaxy NGC 5128. 

Another solution to the G-dwarf problem, one that has received a great deal
of attention recently, is the idea that, at early epochs, gas could have
flowed into the system while star formation was occuring thereby invalidating 
the pure `closed-box' assumption (Larson 1972; Timmes, Woosley, \& Weaver 1995; 
Gibson \& Matteucci 1997). The simplest of these `infall' or `accreting-box' 
models assumes that the gas accretion rate is approximately the same 
as the star formation rate yielding a roughly constant gas mass. In this case,
the metal abundance distribution takes on a hyperbolic form (Harris
\& Harris 2000) given by
\[
\frac{dN}{dZ}\sim \frac{M_g}{y-Z}~~~~~~~~\frac{dN}{d[M/H]} \sim M_g \frac{Z}{y-Z}.
\]
The solid lines in Fig. 8 show the result of fitting this relation to the low
abundance regime of the M31 bulge MDF. From a weighted least squares
fit, we find $y = 0.0059 \pm 0.0002$. Thus the MDF and model comparisons in Fig. 8
suggest that some sort of gas infall occurred early in the history of the M31 bulge 
that provided an initial abundance of $[M/H]$$\sim$--1.6 from which
subsequent generations of stars formed (Wyse \& Gilmore 1993). If
we assume that the M31 bulge stars are enhanced in the $\alpha$-elements,
then we would require a higher yield (y$\sim$0.030) for the closed
box model and a slightly higher
initial abundance ($[M/H]$$\sim$--1.4) to fit the MDF adequately.

We can perform an identical analysis on the Milky Way bulge MDF 
derived from the Z2003 data. The results are shown in Fig. 9 wherein the top
panel shows the MW bulge MDF using [M/H] while the lower panel
displays the $Z/Z_{\odot}$ MDF. The dashed lines show the closed box
fits from Z2003 with $y = $$\langle$$Z$$\rangle$$ = 0.015$
but adopting $Z_0 = 0.0003$ ($[M/H] = -1.8$) instead of $Z_0 = 0.0$. Although 
the number of stars is relatively small in the lower abundance bins, 
the solid lines show a possible accreting box scenario with $y = 0.004$.
This analysis underscores the need to account for the G-dwarf problem
in the MW bulge as well as the M31 bulge. In particular, it suggests
that the accretion of gas into the bulges provides a reasonable process 
by which to account for the G-dwarf problem.

We note that the `gas infall' scenario to explain the MDF of M31's bulge fits nicely
with the results of Durrell, Harris, \& Pritchet (2001). They constructed an MDF
for an M31 halo field at 20 kpc from the center. Based on this and the results
of previous investigators, Durrell et al. (2001) suggest that gas may have 
`leaked' out of the M31 halo and been `accreted' into the bulge. This is precisely 
the solution we have suggested in order to explain the G-dwarf problem
in the M31 bulge MDF. 


Considering the similarities between the MW bulge and that of M31 
along with the apparent  link between the M31 bulge and halo, it is tempting to 
consider the formation and evolution of the spheroids (bulge + halo) of these 
two spiral galaxies within a framework that differs only by a scaling factor.
We can consider each of the spheroids as a unit and explore any
differences between M31 and the MW as simply due to their mass difference. 
To first order, this approach seems reasonable in light of the fact that the
halo stars are likely to be the first to form in the spheroid thus
providing pre-enriched gas for the bulge, which lies at the center of
the gravitational potential. In addition, the discovery of tidal tails
as the remnants of disrupted dwarf galaxy fragments in both the MW
(e.g. Newberg et al. 2002; Majewski et al. 2004) and M31 (e.g.
Ferguson et al. 2002; Ibata et al. 2004) halos suggests that both
structures were formed through the same processes.

Upon further examination, the differences between the M31 and MW halo
stellar populations become quite evident. For example, the peak
metallicity of the Galactic halo is located at [M/H]$\sim$ --1.5 (Ryan
\& Norris 1991) while it is at [M/H]$\sim$ --0.6 for the 20 kpc M31
field of Durrell et al.  (2001). This suggests a fundamental
difference in the values of the yields that shaped the halo MDFs.
Furthermore, the overall shapes of the two halo MDFs are fundamentally
different; the MW distribution is relatively symmetric around the mean
while the M31 MDF exhibits a metal-poor tail. Differences such as
these should not be surprising if the bulk of the halos were built up
through the accretion of disrupted satellite galaxies, since this
process, by its very nature, is rather chaotic. It could thus produce
two spiral halos with the very different properties we see in the MW
and M31.

In light of these points, we are inclined to suggest that
the bulges of the MW and M31 were already mostly 
formed when the halo experienced its most intense epoch of dwarf
galaxy interactions. Otherwise, it is difficult to
understand how the M31 and MW bulges could resemble each other so
much, while embedded in such different halos, whose stellar populations
reveal very different star formation histories.

\section{Conclusions}

We present a VI color-magnitude diagram of a field near the M31 bulge globular
cluster G170 using HST/WFPC2 observations. Using the Girardi et al. (2002)
and Salasnich et al. (2000)
theoretical red giant branches, we have converted the colors of stars to metal
abundance values. These, in turn, are used to constructed a metallicity
distribution function for the bulge of M31. We have analyzed the MDF in both
the [M/H] plane and the $Z/Z_{\odot}$ planes. From this analysis, we draw
the following conclusions.

\noindent 1. The MDF shows a peak at
$[M/H]$$\sim$0 with a steep decline at higher metallicities and a more
gradual tail to lower metallicities. This is similar in shape to the MDF
of the Milky Way bulge but shifted to higher metallicities by $\sim$0.1
dex. 

\noindent 2. A pure closed box model of chemical evolution, even with significant 
pre-enrichment, does not seem to simultaneously match the [M/H] and $Z/Z_{\odot}$
MDFs of M31's bulge. A scenario in which an initial infall of gas enriched the
bulge to an abundance of $[M/H]$$\sim$--1.6 ($[M/H]$$\sim$--1.4
in the $\alpha$-enhanced case) with subsequent evolution
proceeding as a closed box provides a better fit to the observed MDF.

\noindent 3. The significant differences in the MDFs of the M31 and MW halos
are not at all reflected in those of their bulges. Therefore, we
suggest that the bulk of the stars in the bulges of both galaxies were
in place before the accretion events that occurred in the halos could
influence them, and that bulges are formed within a
common framework.

\acknowledgments

We are grateful to Manuela Zoccali for providing the Galactic bulge data in a timely 
manner and to Leo Girardi for his advice regarding the model RGBs. 
We also thank Bill Harris and Mike Barker for enlightening conversations
about the closed box model and its predictions. We acknowledge Aaron Grocholski 
for his help with the IDL code that performed the metallicity interpolation.
An anonymous referee provided valuable comments that significantly improved this
work. This research was supported by NSF CAREER grant AST-0094048.

\clearpage



\begin{figure}
\caption{Color-magnitude diagrams for three M31 bulge regions centered 
on the clusters G170, G198, and G177.}
\end{figure}

\begin{figure}
\caption{Luminosity functions in V and I for the three CMDs shown in Fig. 1.}
\end{figure}

\begin{figure}
\caption{Color-magnitude diagram for the field surrounding G170 adjusted for
distance and reddening (see text). The solid lines are the theoretical RGBs 
of Girardi et al. (2002) for an age of 12.6 Gyr and metallicities of
Z=0.0001, 0.0004, 0.001, 0.004, 0.008, and 0.019. The two reddest RGBs
are taken from Salasnich et al. (2000) for metallicities of Z=0.04 and 0.07. }
\end{figure}

\begin{figure}
\caption{The metallicity distribution function for the G170 bulge field in the
[M/H] plane (upper panel) and in the $Z/Z_{\odot}$ plane (lower panel). The
solid points represent the binned histograms while the dashed lines are
the generalized histograms.}
\end{figure}

\begin{figure}
\caption{(a) The solid lines are the theoretical RGBs 
of Girardi et al. (2002) for an age of 12.6 Gyr and metallicities of
Z=0.0001, 0.0004, 0.001, 0.004, 0.008, and 0.019. The two reddest RGBs
are taken from Salasnich et al. (2000) for metallicities of Z=0.04 and 0.07. 
The plotted points are Milky Way bulge giants with V--K photometry from
Zoccali et al. (2003). The filled circles represent first ascent red giant stars
used in conjunction with the plotted theoretical RGB sequences to derive the
metallicity distribution function (MDF) of the Milky Way bulge shown in (b). 
The open circles are 
asymptotic giant branch stars and are excluded from the MDF sample.
(b) The MDF for the Milky Way bulge derived from the data in (a). }
\end{figure}

\begin{figure}
\caption{ The metallicity distribution function for the G170 bulge field
(filled circles and dashed line) compared with that of the Milky Way
bulge derived using the V--K photometry of Zoccali et al. (2003) coupled
with the model RGBs from Girardi et al. (2002) and Salasnich et al. (2000)
as shown in Fig. 5. Both distributions have been scaled to unit area.}
\end{figure}

\begin{figure}
\caption{The metallicity distribution function for the G170 bulge field
(filled circles) compared with the predictions of a simple closed box model
(dashed line) with zero pre-enrichment ($Z_0$ = 0.0) and 
$y=$$\langle$$Z$$\rangle$=0.019.}
\end{figure}

\begin{figure}
\caption{The metallicity distribution function for the G170 bulge field
(filled circles) compared with the predictions of two chemical enrichment models.
The dashed line is the closed box model with pre-enrichment 
amounting to $Z_0$ = 0.0005 ($[M/H] = -1.6$) and 
$y=$$\langle$$Z$$\rangle$=0.019. The solid line is an
`accreting' model with $y=0.0059$.}
\end{figure}

\begin{figure}
\caption{Same as Fig. 8 except that the metallicity distribution function for the 
Milky Way bulge is plotted using the data of Zoccali et al. (2003). The dashed line 
is the closed box model with pre-enrichment of $Z_0$ = 0.0003 ($[M/H] = -1.8$) and 
$y=$$\langle$$Z$$\rangle$=0.015. The solid line is a possible `accreting' model 
that fits the low Z end of the MDF.}
\end{figure}

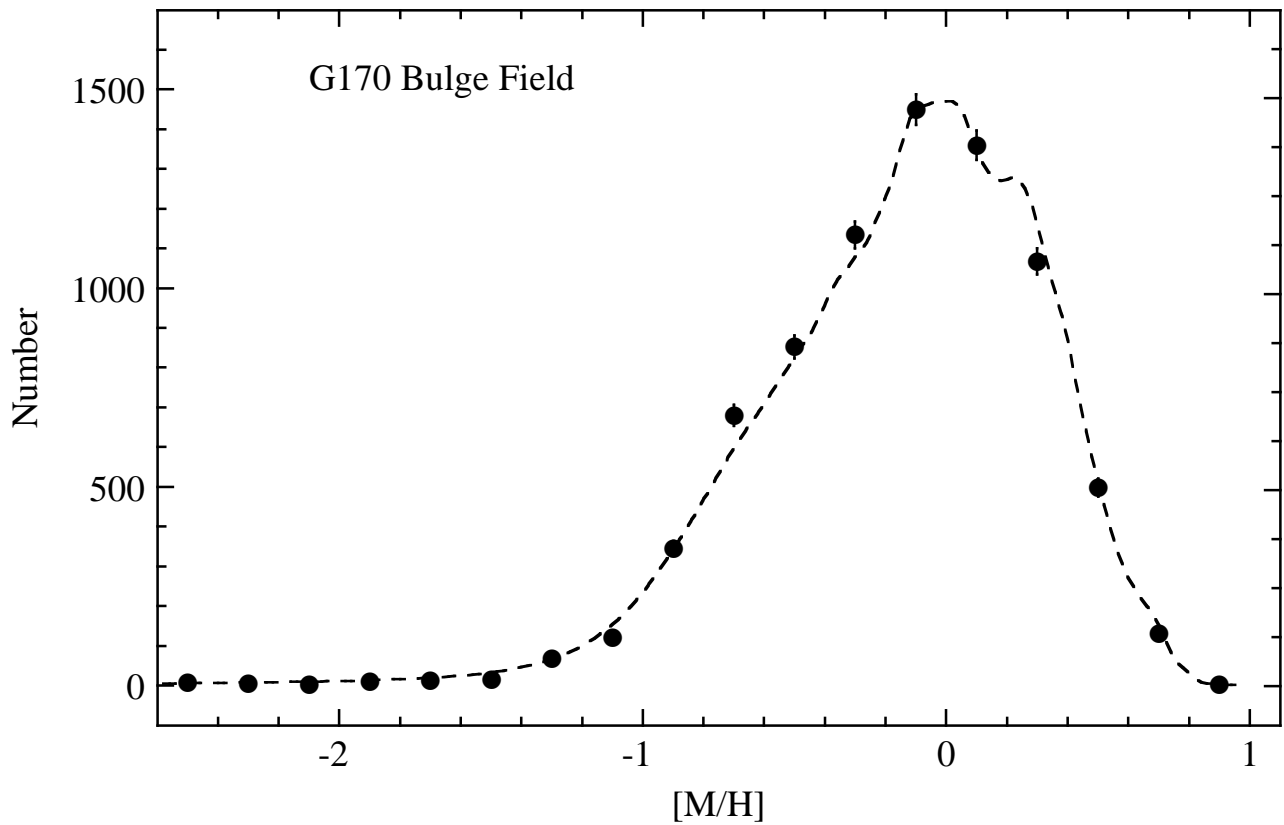
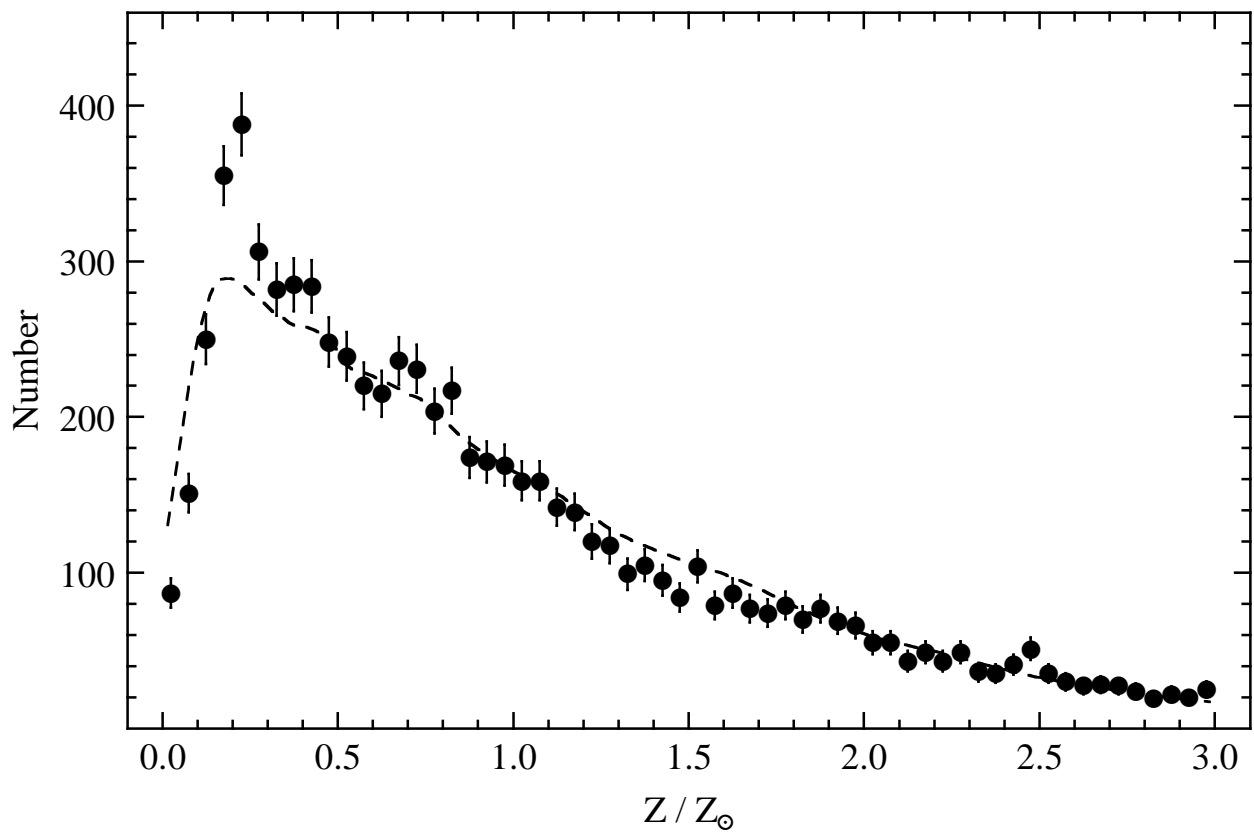
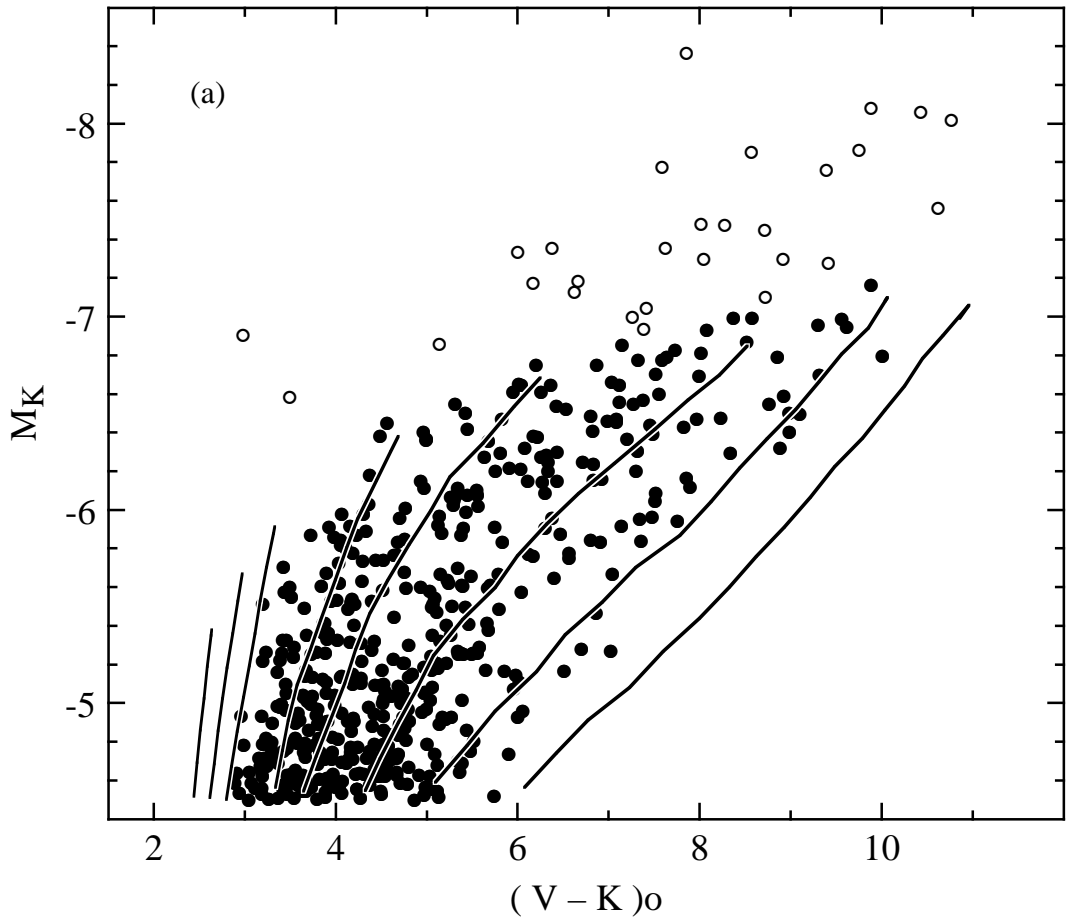
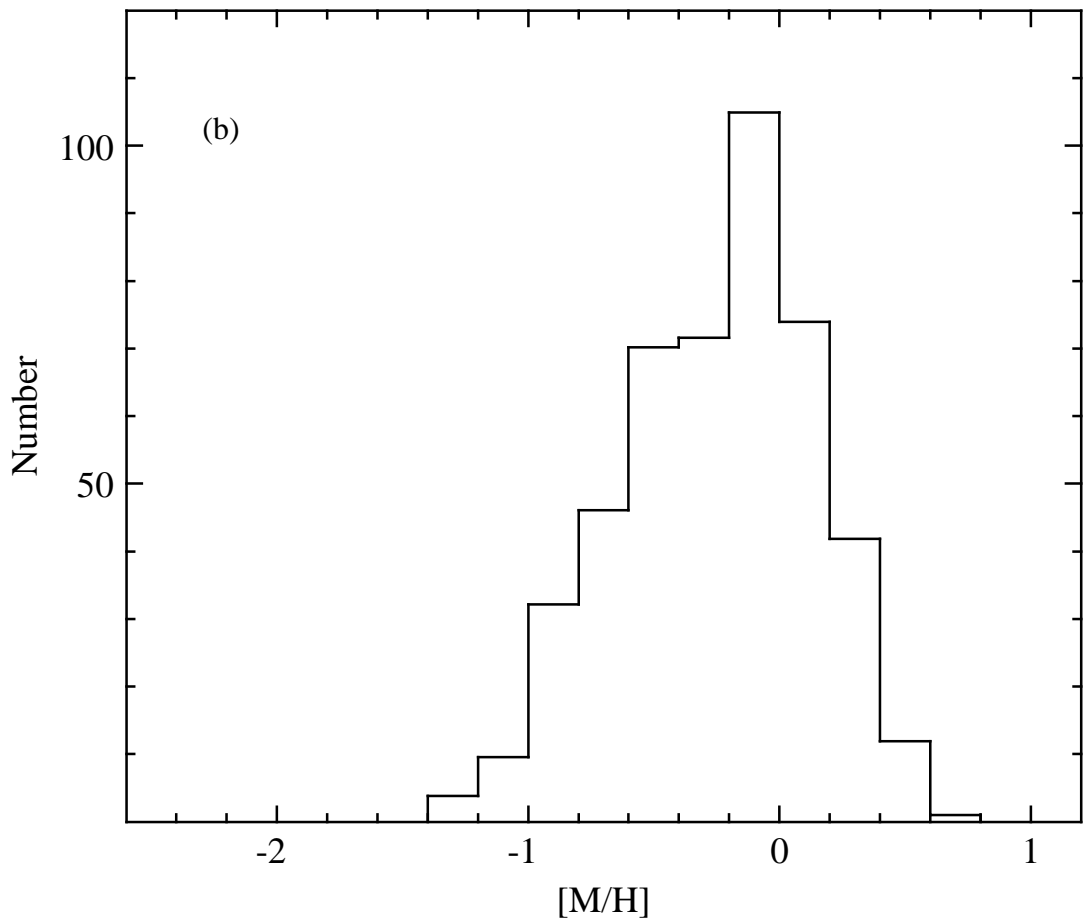
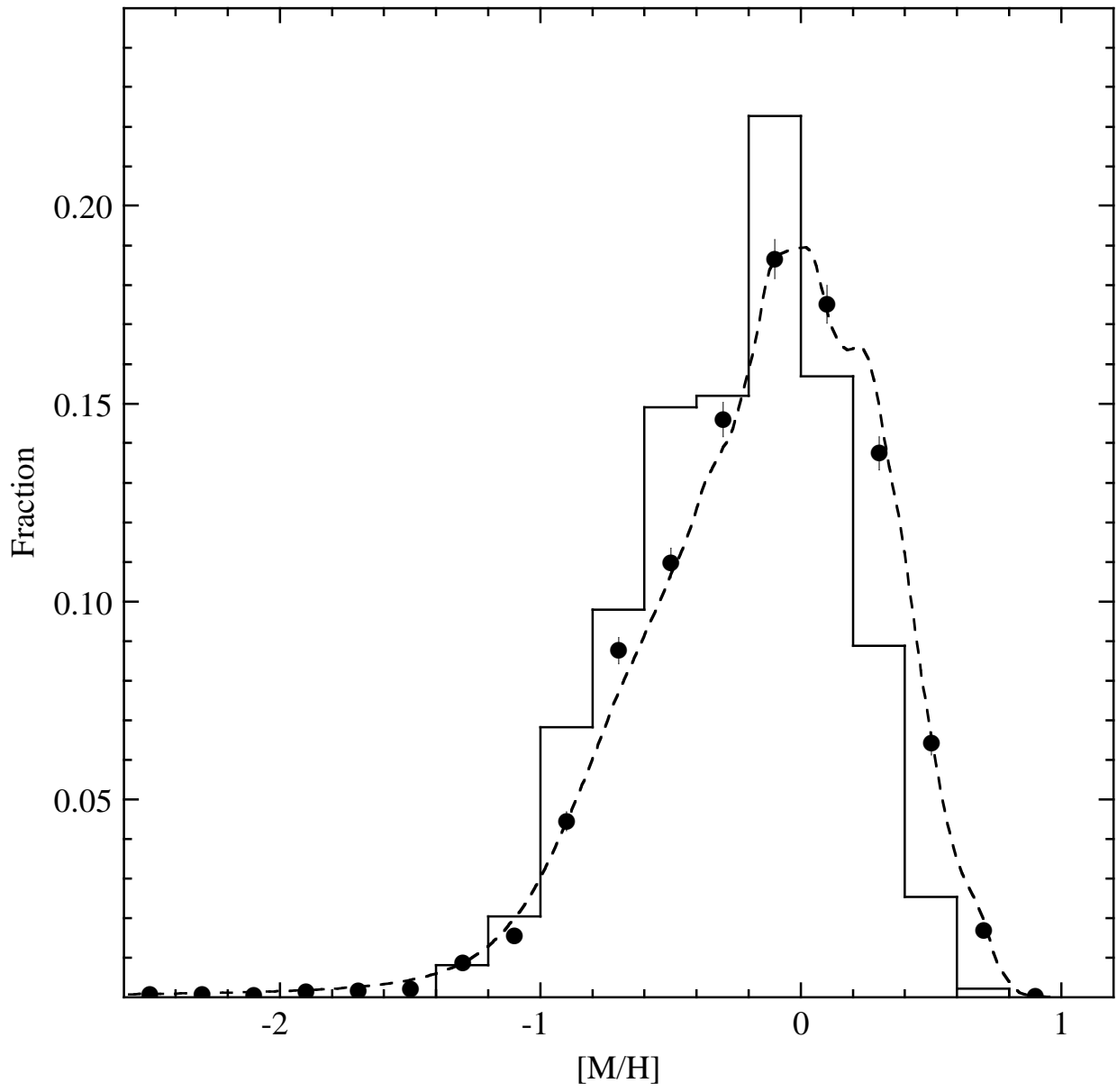
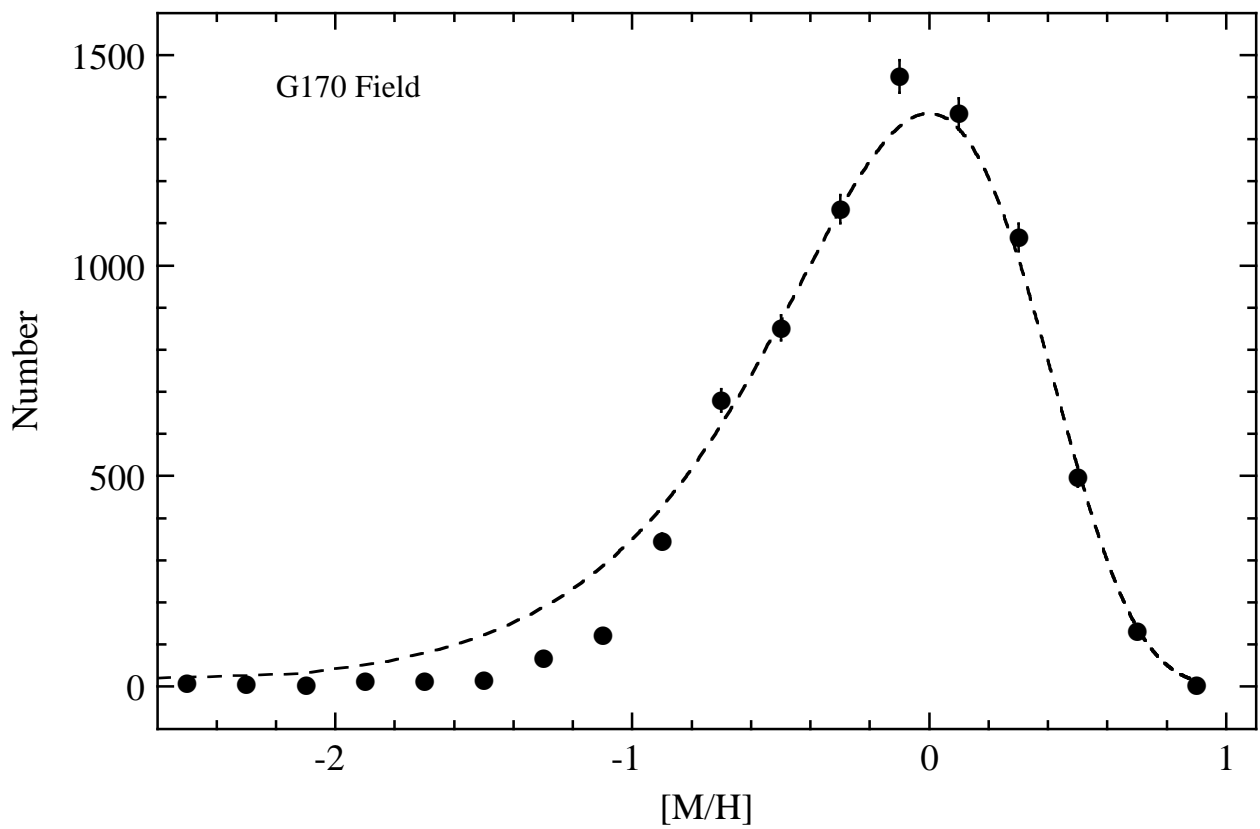
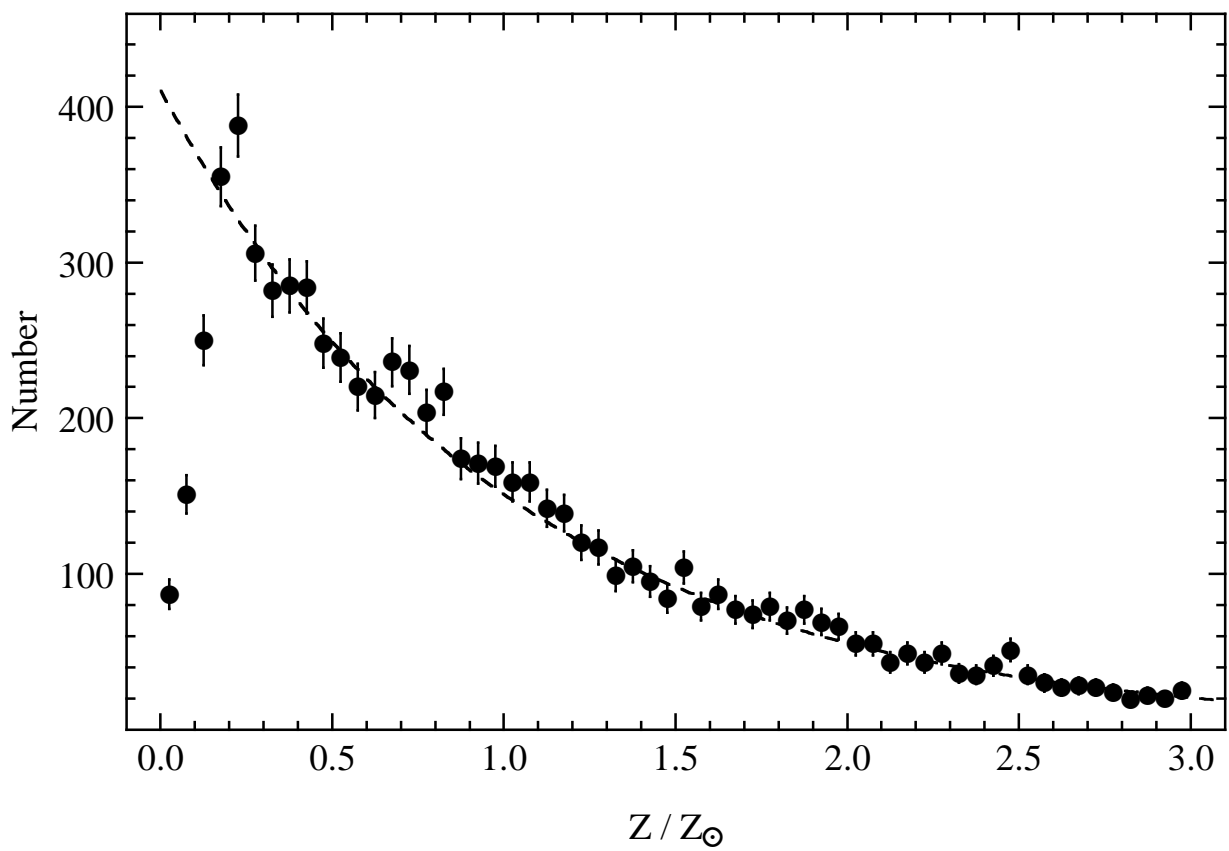
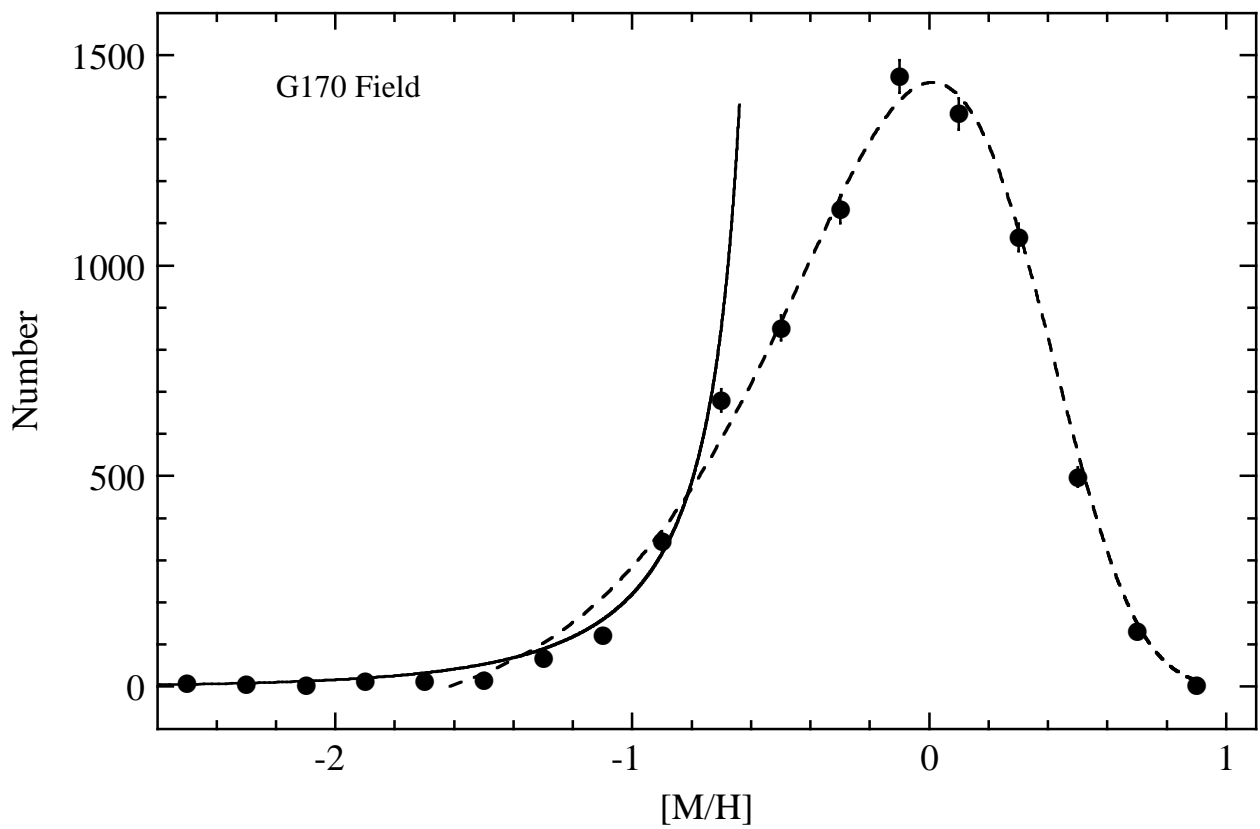
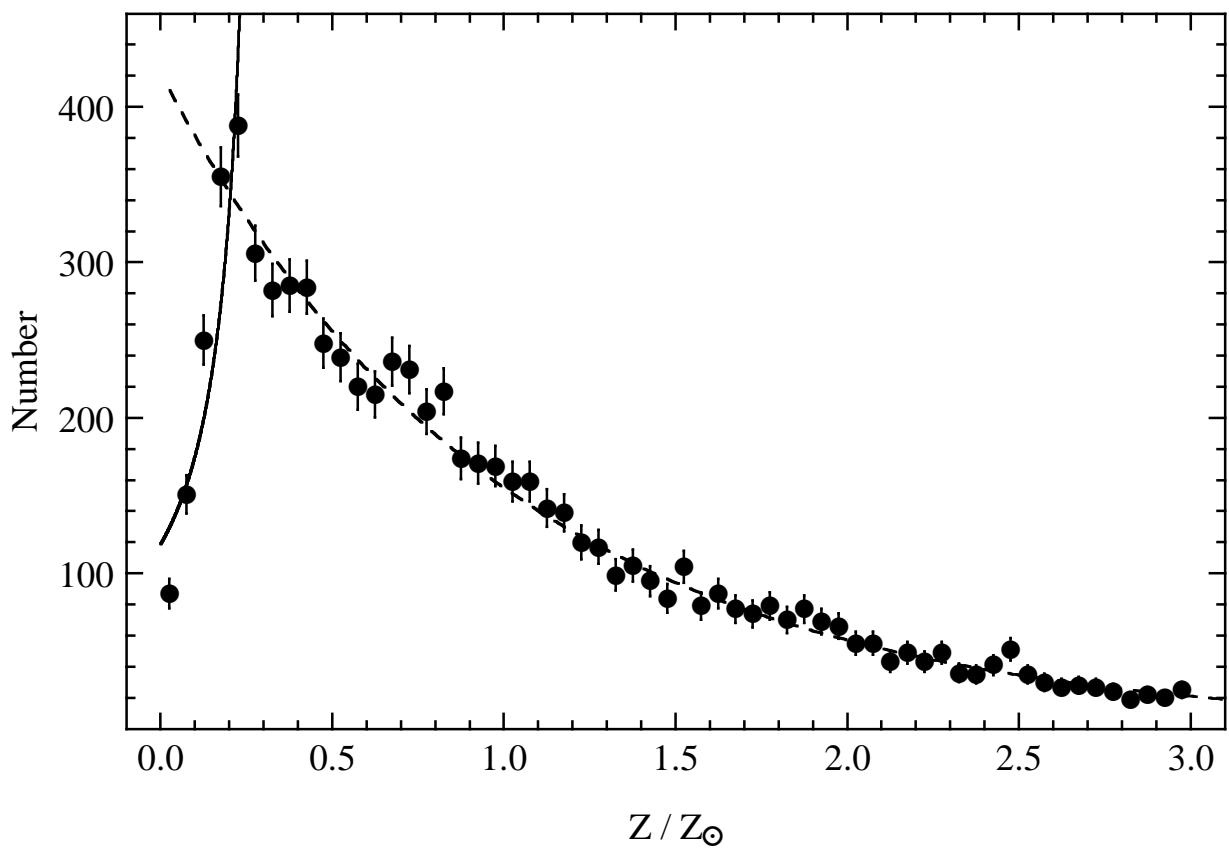
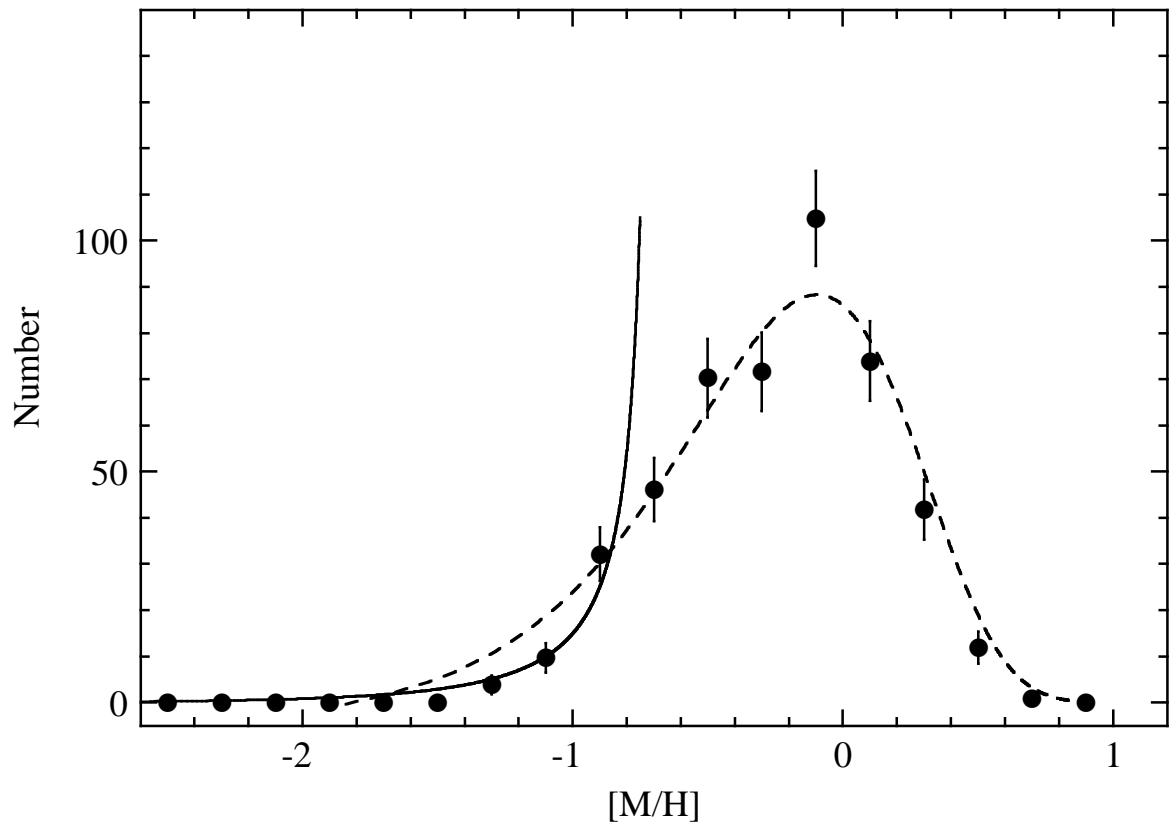
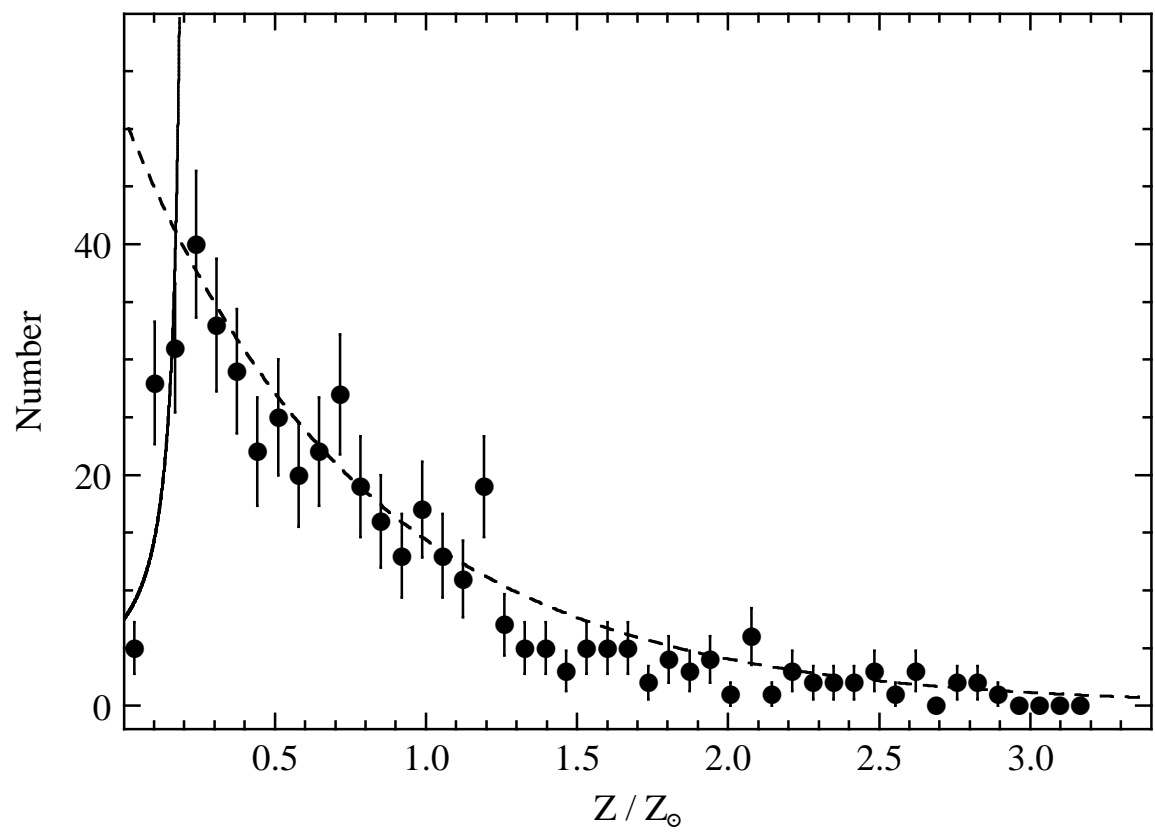
\end{document}